\documentclass[submission,copyright,creativecommons]{eptcs}
\providecommand{\event}{TTATT 2013}

\usepackage[utf8]{inputenc}
\usepackage[T1]{fontenc}
\usepackage{graphicx}
\usepackage{amsmath,amssymb,amsthm}
\usepackage{enumerate}
\usepackage{accents}
\hypersetup{%
 bookmarksnumbered=true,%
 setpagesize=false,%
 colorlinks=false,%
 linkbordercolor={0.8 0.8 0.8},%
 citebordercolor={0.8 0.8 0.8},%
 pdfborder={0 0 0.6},%
 pdftitle={Node Query Preservation for Deterministic Linear Top-down Tree Transducers},%
 pdfauthor={Kazuki Miyahara, Kenji Hashimoto, Hiroyuki Seki},%
 pdfsubject={TTATT 2013},%
 pdfkeywords={XML, tree automata, tree transducers, run-based queries, query preservation} %
}
\usepackage{pict2e}
\usepackage{qtree}

\allowdisplaybreaks[3]

\makeatletter
  \def\widebar{\accentset{{\cc@style\underline{\mskip10mu}}}}
\makeatother

\newcounter{stepWQP}
\newcounter{stepWQC}
\newcounter{stepWQPN}
\newcounter{stepQP}
\newcommand{\addStepWQP}[1][]{\refstepcounter{stepWQP}{#1}Step~{\thestepWQP}.}
\newcommand{\addStepWQC}[1][]{\refstepcounter{stepWQC}{#1}Step~{\thestepWQC}.}
\newcommand{\addStepWQPN}[1][]{\refstepcounter{stepWQPN}{#1}Step~{\thestepWQPN}.}

\newcommand{\addStepQP}[1][]{\refstepcounter{stepQP}{#1}Step~{\thestepQP}.}

\theoremstyle{plain}
\newtheorem{theorem}{Theorem}
\newtheorem{lemma}{Lemma}

\theoremstyle{definition}

\newtheorem{example}{Example}

\newcommand{\figlab}[1]{\label{fig:#1}}
\newcommand{\figref}[1]{Figure~\ref{fig:#1}} 
\newcommand{\forlab}[1]{\label{for:#1}}
\newcommand{\forref}[1]{(\ref{for:#1})} 
\newcommand{\seclab}[1]{\label{sec:#1}}
\newcommand{\secref}[1]{\ref{sec:#1}} 

\newcommand{\lmmlab}[1]{\label{lmm:#1}}
\newcommand{\thmlab}[1]{\label{thm:#1}}

\newcommand{\steplab}[1]{\label{step:#1}}
\newcommand{\stepref}[1]{step~\ref{step:#1}} 
\newcommand{\Stepref}[1]{Step~\ref{step:#1}}

\newcommand{\trel}[1]{[\![#1 ]\!]}
\newcommand{\lab}{\operatorname{lab}}
\newcommand{\val}{\operatorname{val}}
\newcommand{\dom}{\operatorname{dom}}
\newcommand{\rng}{\operatorname{rng}}
\newcommand{\RUN}{\operatorname{run}}
\newcommand{\stat}{\operatorname{state}}
\newcommand{\rk}{\operatorname{rk}}
\newcommand{\tr}{T\!r}

\newcommand{\LQ}{L_Q}
\newcommand{\LT}{L_T}
\newcommand{\Lin}{\mathrm{L}}
\newcommand{\Det}{\mathrm{D}}
\newcommand{\TT}{\mathrm{T}}
\newcommand{\LTT}{\Lin\TT}
\newcommand{\DLTT}{\Det\Lin\TT}
\newcommand{\RQ}{1\mathchar`-\mathrm{RQ}}
\newcommand{\RQN}{n\mathchar`-\mathrm{RQ}}
\newcommand{\WQP}{1\mathchar`-\mathbf{WQP}}

\newcommand{\WQPN}{n\mathchar`-\mathbf{WQP}}
\newcommand{\QPN}{n\mathchar`-\mathbf{QP}}

\newcommand{\WQC}{1\mathchar`-\mathbf{WQC}}
\newcommand{\WQCN}{n\mathchar`-\mathbf{WQC}}

\newcommand{\PD}{P^\mathit{del}}
\newcommand{\TS}{T_\Sigma}
\newcommand{\TD}{T_\Delta}
\newcommand{\TSN}{\TS^{(\mathbb{N})}}
\newcommand{\TDN}{\TD^{(\mathbb{N})}}
\newcommand{\mk}{\mathrm{mk}}
\newcommand{\idx}{\mathrm{idx}}
\newcommand{\pos}{\operatorname{pos}}
\newcommand{\ctx}{\mathbf{C}}
\newcommand{\vset}{X}

\title{Node Query Preservation for Deterministic Linear Top-Down Tree Transducers}

\author{Kazuki Miyahara
\institute{Nara Institute of Science and Technology\\[0.5mm]Nara, Japan}
\email{kazuki-mi@is.naist.jp}
\and
Kenji Hashimoto \qquad\qquad Hiroyuki Seki
\institute{Nagoya University\\[0.5mm]Nagoya, Japan}
\email{\quad \{k-hasimt,seki\}@is.nagoya-u.ac.jp}
}
\def\titlerunning{Node Query Preservation for Deterministic Linear Top-Down Tree Transducers}
\def\authorrunning{K.~Miyahara, K.~Hashimoto \& H.~Seki}

\begin{document}

\maketitle

\begin{abstract}
This paper discusses the decidability of node query preservation problems for XML document transformations.
We assume a transformation given by a deterministic linear top-down data tree transducer
(abbreviated as~$\DLTT^V$)
and an $n$-ary query based on runs of a tree automaton.
We say that a $\DLTT^V~\tr$ strongly preserves a query~$Q$ if there is a query~$Q'$ such that
for every document~$t$, the answer set of~$Q'$ for~$\tr(t)$ is equal to
the answer set of~$Q$ for~$t$.
Also we say that $\tr$ weakly preserves~$Q$ if there is a query~$Q'$ such that
for every~$t_d$ in the range of~$\tr$, the answer set of~$Q'$ for~$t_d$ is equal to
the union of the answer set of~$Q$ for~$t$ such that $t_d=\tr(t)$.
We show that the weak preservation problem is coNP-complete and the strong preservation problem is in~2-EXPTIME\null.
\end{abstract}

\section{Introduction}
\label{sect:introduction}
Due to data exchanges and schema updates, long-term databases often require
XML document transformations.
A fundamental concern with these XML document transformations is that the information
contained in each source document should be preserved in the target document obtained
by transforming the source document.
Query preservation~\cite{Bohannon05} is one of the formulations of such information preservation.
A transformation preserves a query~$Q$ over the source documents if there is a query~$Q'$
on the target documents such that the answer of~$Q$ on a source document
is equivalent to the answer of~$Q'$ on the target document. 

The query preservation problem was shown in~\cite{Bohannon05} to be undecidable for the class
of transformations and queries which can simulate first-order logic queries and projection queries, respectively, in the relational data model. In~\cite{Benedikt13,Hashimoto13}, query preservation has been studied in a setting such that
transformations and queries are both modeled by tree transducers.
In~\cite{Hashimoto13}, the query preservation was shown to be decidable for compositions of
functional linear extended top-down tree transducers (with regular look-ahead)
as transformations, and for either deterministic top-down tree transducers
(with regular look-ahead) or deterministic MSO definable tree
transducers as queries.
Also, query preservation (or, determinacy) was considered in~\cite{Groz13}
as relations between XML views in the context of unranked trees.
In~\cite{Groz13}, views are defined as transformations that leave nodes selected
by node queries, such as Regular XPath queries and MSO queries.
In this paper, we focus on query preservation of a tree transducer for
a node query model,
that is, types of models for transformations and queries are different.
We consider $n$-ary queries, which extract sets of $n$-tuples of nodes.
Query preservation for tree-to-tree queries requires that the tree
structure of the
source query result should be restored by some query to the transformed data.
In our setting, query preservation requires that a tree transducer
should maintain
the possibility to extract the relationship between (the values of)
the nodes specified by
a node query, rather than the tree structures.

In this paper, we model an XML document by a data tree, which is
 a ranked ordered tree
where each node can have any nonnegative integer as a data value.
We assume a transformation given by a deterministic linear top-down data tree transducer
(abbreviated as~$\DLTT^V$) and a run-based $n$-ary query~\cite{Niehren05}
(equivalent with an MSO $n$-ary query~\cite{Thatcher68}).
As defined in the next section, the transformation is determined independently of data values assigned to nodes,
though some data values can be transferred from input to output without duplication.
The answer set of a query is the set of tuples of data values
which are assigned to nodes selected by the query
instead of the selected nodes themselves.
We say that a $\DLTT^V~\tr$ strongly preserves a query~$Q$ if there is a query~$Q'$ such that
for every document~$t$, the answer set of~$Q'$ for~$\tr(t)$ is equal to
the answer set of~$Q$ for~$t$.
Also we say that $\tr$ weakly preserves~$Q$ if there is a query~$Q'$ such that
for every~$t_d$ in the range of~$\tr$, the answer set of~$Q'$ for~$t_d$ is equal to
the union of the answer set of~$Q$ for~$t$ such that $t_d=\tr(t)$.
We show that the weak query preservation problem is coNP-complete.
If the tuple size~$n$ of queries is a constant, the complexity becomes PTIME\null.
We also show that the strong preservation problem is in~2-EXPTIME\null.
The decidability result of the two cases can be extended to the situation where the transformation is
given by a~$\DLTT^V$ with regular look-ahead.

\section{Preliminaries}
\label{sect:typesetting}
\subsection{Data Trees}

We denote the set of all nonnegative integers by~$\mathbb{N}$.
For~$n\in\mathbb{N}$, the set~$\{1,\dots,n\}$ is denoted by~$[n]$.
A (ranked) alphabet is a finite set~$\Sigma$ of symbols with a mapping
$\rk : \Sigma \to \mathbb{N}$.
Let
$\Sigma_n = \{\,\sigma\in\Sigma \mid \rk(\sigma)=n\,\}$.
A {\itshape data tree\/} is a tree such that each symbol of the tree
can have a nonnegative integer as a data value.
Formally, the set~$\TSN$ of data trees over an alphabet~$\Sigma$ is
the smallest set~$T$ such that $\sigma(t_1,\dots,t_n)\in T$~and~$\sigma^{(\nu)}(t_1,\dots,t_n) \in T$
for every $\sigma\in\Sigma_n$, $t_1,\dots,t_n\in T$~and~$\nu\in\mathbb{N}$.
For a data tree~$t$, the set of positions (nodes)~$\pos(t)$ is defined in the usual way and
let $t/v$ denote the subtree of~$t$ at position~$v\in\pos(t)$.
If $t/v = \sigma^{(\nu)}(t_1,\dots,t_n)$, we write $\lab(t,v)=\sigma$~and~$\val(t,v)=\nu$.
If $t/v=\sigma(t_1,\dots,t_n)$, we write $\lab(t,v)=\sigma$.
Let $t[v\leftarrow t']$ be the tree obtained from~$t$ by replacing $t/v$~with~$t'$.
A data tree~$t$ is {\itshape proper\/} if every symbol appearing in~$t$ has a value.
A {\itshape tree\/} is a data tree that does not contain any value.
Let $\TS$ denote the set of all trees over~$\Sigma$.
For a data tree~$t$, let $t^-$ denote the tree obtained from~$t$ by removing
all the values in~$t$.
For every~$n\geq 1$, let $\vset_n=\{\,x_i \mid i \in [n]\,\}$
be a set of variables with $\rk(x_i)=0$ for every $x_i \in \vset_n$.
A tree~$t$ is {\itshape linear\/} if each variable occurs at most once in~$t$.
A linear tree in~$T_{\Sigma\cup\vset_n}$ is called an ($n$-ary) {\itshape context\/}
over~$\Sigma$.
Let $\ctx(\Sigma,\vset_n)$ denote the set of $n$-ary contexts over~$\Sigma$.
For a context~$C\in\ctx(\Sigma, \vset_n)$, let $C[t_1,\dots,t_n]$
denote the tree obtained from~$C$ by replacing $x_i$~with~$t_i$ for~$1\le i\le n$.

\subsection{Tree Automata and Tree Transducers}

A {\itshape tree automaton\/} (TA) is a tuple $A=(P, \Sigma, P_I, \delta)$
where $P$ is a finite set of states, $\Sigma$ is a ranked alphabet,
$P_I\subseteq P$ is a set of initial states,
and $\delta$ is a finite set of transition rules of the form
\[
 p \to \sigma(p_1,\dots,p_n)
\]
where $p\in P$, $\sigma\in\Sigma_n$, and $p_1,\dots,p_n\in P$.
Let~$\stat(A) = P$.
TA~$A$ accepts a tree~$t\in\TS$ if there is a mapping
$m: \pos(t)\to P$ such that (1)~$m(\varepsilon) \in P_I$, and (2)~for $v\in\pos(t)$
with $t/v=\sigma(t_1,\dots,t_n)$, $m(v)\to\sigma(m(v1),\dots,m(vn))\in\delta$.
The mapping~$m$ is called an {\itshape accepting run\/} of~$A$ on~$t$.
The set of all accepting runs of~$A$ on~$t$ is denoted by~$\RUN(A,t)$.
Let~$L(A) = \{\, t\in\TS \mid \RUN(A,t) \neq \emptyset \,\}$.
We simply write a run to mean an accepting run.
A state of~$A$ is {\itshape useless\/} if it is not assigned to any position by any accepting run of~$A$,
and a rule is useless if it has a useless state.
A TA~$A$ is said to be {\itshape reduced}, if $A$ has no useless states and transition rules.


A {\itshape linear top-down data tree transducer\/} $(\LTT^V)$ is a tuple
$\tr=(P, \Sigma, \Delta, P_I, \delta)$ where $P$ is a finite set of states,
$\Sigma$~and~$\Delta$ are ranked alphabets of input and output, respectively,
$P_I \subseteq P$ is a set of initial states, and $\delta$ is a finite set of transduction rules of the form
\[
p(\sigma^{(z)}(x_1,\dots,x_n))\ \to\ C^{(j\gets z)}[\,p_1(x_1), \dots, p_n(x_n)\,],
\]
 where $p, p_1, \dots, p_n\in P$, $\sigma \in \Sigma_n$,
 $j\in\{\, v \mid v\in\pos(V), t/v \notin\vset_n \,\}$,
 $x_1, \dots, x_n \in \vset_n$, $C\in\ctx(\Delta,\vset_n)$,
 and $(j\gets z)$ is optional.
We call $(j \gets z)$ the {\itshape value position designation\/} of the rule.
The move relation~$\Rightarrow_{\tr}$ of an $\LTT^V~\tr=(P,\Sigma,\Delta,P_I,\delta)$ is defined as follows:
If $p(\sigma^{(z)} (x_1,\dots,x_n)) \to C^{(j\gets z)}[\,p_1(x_1),\dots,p_n(x_n)\,] \in \delta$,
$t_1,\dots,t_n\in\TSN$ and $t/v=p(\sigma^{(\nu)}(t_1,\dots,t_n))$ $(\nu\in\mathbb{N})$,
then
\[
 t \Rightarrow t\,[\,v\gets C^{(j\gets \nu)}[\,p_1(t_1),\dots,p_n(t_n)]\,]
\]
where $C^{(j\gets \nu)}$ is the context obtained from~$C$ by replacing $\lab(C,j)$ with $\lab(C,j)^{(\nu)}$.
When the value position designation is missing in the rule, we do not copy $\nu$ to any position
of the output.
Let $\trel{\tr} = \{\,(t, t') \mid p_I(t)\Rightarrow_{\tr}^* t',\,t\in\TSN, t\text{ is proper}, t' \in \TDN, \, p_I \in P_I \,\}$. The domain of~$\tr$ is defined as
$\dom(\tr) = \{\, t \mid \exists t'.\ (t, t') \in \trel{\tr} \,\}$,
and the range of~$\tr$ is defined as
$\rng(\tr)=\{\, t' \mid \exists t.\ (t, t') \in \trel{\tr}\,\}$.
An $\LTT^V~\tr=(P,\Sigma,\Delta,P_I,\delta)$ is {\itshape deterministic\/}
(denoted as $\DLTT^V$) if
(1)~$\left|P_I\right|=1$, and
(2)~for each $p \in P$~and~$\sigma\in\Sigma$, there exists at most one transduction rule that contains
both $p$~and~$\sigma$ in its left-hand side.
If $\tr$ is deterministic,
there is only one pair~$(t, t')\in \trel{\tr}$ for each~$t\in\dom(\tr)$.
Thus, we write~$\tr(t)=t'$ when $(t, t')\in \trel{\tr}$.
For $L\subseteq\TSN$, we write $\tr(L)=\{\, \tr(t) \mid t\in L \,\}$.
We denote by $\tr^{-1}$ the inverse of $\tr$, i.e., $\tr^{-1}(t')=\{\,t\mid \tr(t)=t'\,\}$.
Let $\DLTT$ be the class of ordinary deterministic linear top-down tree transducers
over trees containing no data values.

A {\itshape subtree-deleting\/} rule is a rule such that
at least one variable in its left-hand side does not occur in its right-hand side
as $p_1(\sigma^{(z)}(x_1,x_2)) \to \sigma'^{(z)}(p_2(x_2))$.
A {\itshape value-erasing\/} rule is a rule that does not have the value position designation
in its right-hand side.

\subsection{Run-based {\large $n$}-ary Queries}
A {\itshape run-based $n$-ary query\/} ($\RQN$)~\cite{Niehren05} is a pair~$(A,S)$
where $A=(P,\Sigma,P_I,\delta)$ is a TA and $S\subseteq P^n$.
In this paper, we assume that each $s\in S$ consists of $n$ different states.
We simply call a run-based $n$-ary query a query.
For a data tree~$t$ and a query~$Q=(A,S)$, define
\[
 Q(t) = \bigcup_{m\in \RUN(A,t^-)}  Q(m,t),
\]
where $Q(m,t) = \{\, (v_1,\dots,v_n) \mid (m(v_1),\dots, m(v_n))\in S,\ v_1,\dots,v_n \in \pos(t) \,\}$.
For an $\RQN~Q$ and a data tree~$t$, let
$\val(Q(t))=\{\,(\val(t,v_1),\dots,\val(t,v_n)) \mid (v_1,\dots,v_n)\in Q(t)\,\}$.

We assume that for all query~$Q=(A,S)$, the TA~$A$ is reduced.
\vspace{3mm}
\begin{example}
 Consider $2\mathchar`-\mathrm{RQ}\ Q=(A,S)$ defined by:
 $A=(P, \Sigma, P_I, \delta)$, $P=\{p_1, p_2, p_3, p_4\}$, $\Sigma_2=\{f\}$,
$\Sigma_0=\{a\}$, $P_I=\{p_1\}$,
$\delta=\{\,
 p_1 \to f(p_2,p_3),
 p_2 \to a,\
 p_3 \to f(p_4,p_3),\
 p_3 \to a,\
 p_4 \to a\,\}$, and $S=\{(p_2,p_3)\}$.
 \figref{2rq} shows that the result of the query on the data tree
 \[
 t=f^{(1)}(a^{(2)},f^{(3)}(a^{(4)},a^{(5)}))
 \]
 is $\val(Q(t))=\{(2,3),(2,5)\}$, where the numbers 1 to 5 are the data values on~$t$.
 \begin{figure}[ht]
  \begin{center}
   \includegraphics[width=250pt]{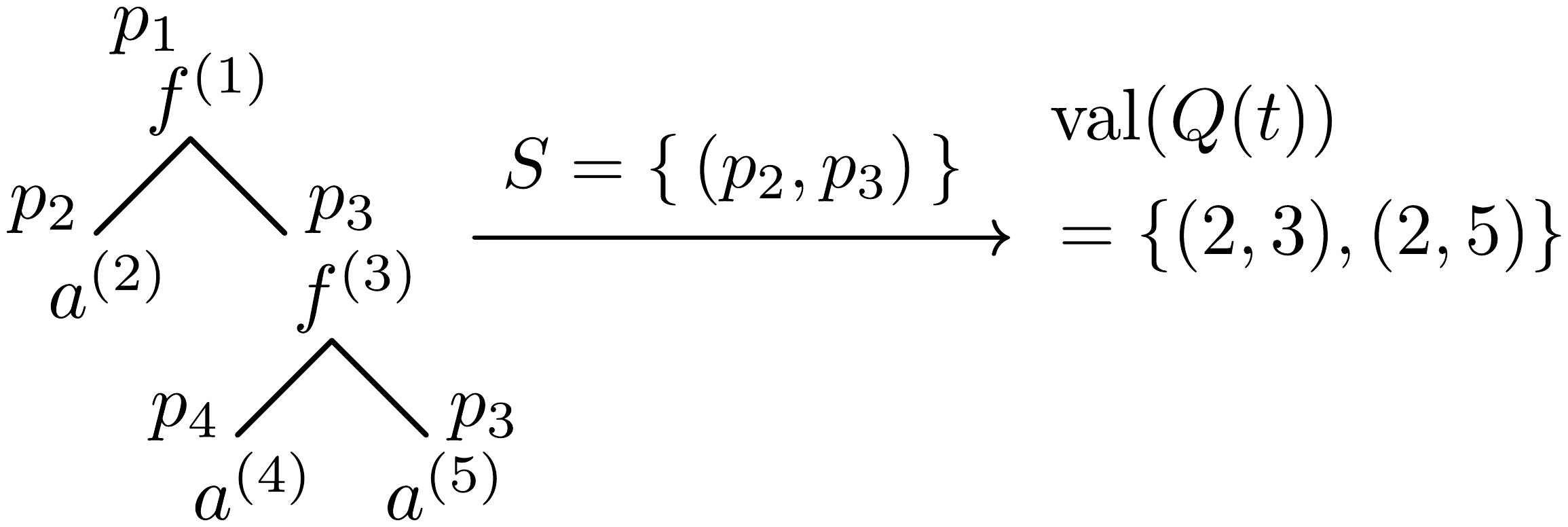}
  \end{center}
  \vspace{-3mm}
  \caption{Example of $2\mathchar`-\mathrm{RQ}$.}
  \figlab{2rq}
 \end{figure}
\end{example}

\subsection{Query Preservation}
Let $\LT$~and~$\LQ$ be a class of tree transducers and a class of queries, respectively.
Given a query~$Q\in\LQ$ and a tree transducer~$\tr\in\LT$,
we say that $\tr$ {\itshape (strongly) preserves\/}~$Q$ if there exists~$Q'\in \LQ$ that satisfies
\begin{equation}
 \val(Q'(\tr(t))) = \val(Q(t)) \forlab{sqp}
\end{equation}
for any~$t\in\dom(\tr)$.
We also define the {\itshape weak\/} query preservation to simplify the discussions about
 the strong query preservation and its decision algorithm (see Section~\secref{SQP}).
We say that the transducer~$\tr$ {\itshape weakly preserves\/} the query~$Q$
if there exists~$Q'\in \LQ$ such that
for any $t_d \in \rng(\tr)$,
\begin{equation}
 \val(Q'(t_d)) = \bigcup_{t\in\tr^{-1}(t_d)} \val(Q(t)). \forlab{qp}
\end{equation}
By definition, we can see that $\tr$ preserves $Q$ if and only if
(1)~$\tr$ weakly preserves $Q$ and (2)~for each $t_d\in \rng(\tr)$ and
any two $t_1,t_2\in \tr^{-1}(t_d)$, $\val(Q(t_1))=\val(Q(t_2))$ holds.

\begin{example}
\label{ex:1}
Let $Q=(A,\{p_1\})$ where $A=(\{p_0,p_1,p_2\},\allowbreak \{ f,g,a\}, \{p_0\},\allowbreak
\{p_0\to f(p_1,p_2),\ p_0\to g(p_2,p_1),\allowbreak\ p_1\to a,\allowbreak\ p_2\to a\})$.
Let $\tr$ be a $\DLTT^V$ defined by the homomorphism that maps
$f,g,a$ to $h,h,a$, respectively (and moves each data value as well).
We can see that $L(A)=\{f(a,a), g(a,a)\}$.
Let $t_1 = f^{(3)}(a^{(4)},a^{(5)})$ and $t_2 = g^{(3)}(a^{(4)},a^{(5)})$.
 \begin{figure}[ht]
  \begin{center}
   \qtreecenterfalse
   Tree~$t_1$. \Tree [.$f^{(3)}$ \\[-2mm]$a^{(4)}$ !\qsetw{1.5cm} \\[-2mm]$a^{(5)}$ ]
   \hspace{5mm}
   Tree~$t_2$. \Tree [.$g^{(3)}$ \\[-2mm]$a^{(4)}$ !\qsetw{1.5cm} \\[-2mm]$a^{(5)}$ ]
  \end{center}
  \figlab{example2}
  \vspace{-3mm}
 \end{figure}
Then $\tr(t_1)=\tr(t_2)=h^{(3)}(a^{(4)},a^{(5)})$.
In this example, $\tr$ weakly preserves~$Q$.
In fact, $Q'$ obtained from~$Q$ by replacing the first two rules
of~$A$ with $p_0\to h(p_1,p_2), p_0\to h(p_2,p_1)$ satisfies
Equation~\forref{qp}.
On the other hand, $\tr$ does not preserve~$Q$ because
$\val(Q(t_1)) = \{4\} \neq \{5\} = \val(Q(t_2))$
while
$\tr(t_1)=\tr(t_2)$, which imply that
there is no $Q'$ that satisfies Equation~\forref{sqp}.
\end{example}

\section{Decidability of Weak Query Preservation}
\subsection{Unary Queries}
\subsubsection{Decidability}
We show an algorithm that decides if a~$\DLTT^V~\tr$ weakly preserves a $\RQ~Q$.
We also prove that if $\tr$~weakly preserves~$Q$,
we can effectively construct a $\RQ~Q'$ that satisfies Equation~\forref{qp}
(the condition of the weak query preservation).
Assume $Q=(A,\{p\})$ with $p\in \stat(A)$ for a while.
Our algorithm for weak query preservation decides
if there exist a tree~$t\in\dom(\tr)$ and a position~$\widetilde{v}\in\pos(t)$
satisfying the next conditions:
\begin{itemize}
 \item There exists a run $m \in \RUN(A,t^-)$ such that $m(\widetilde{v})=p$.
 \item The data load at~$\widetilde{v}$ on~$t$ is ``removed''
 by a subtree-deleting rule or a value-erasing rule of~$\tr$.
\end{itemize}
Assume there exist a data tree~$t$ and a position~$\widetilde{v}$ of~$t$ that satisfy the above conditions.
Let~$t_d=\tr(t)$, then,
\[
  \bigcup_{t'\in\tr^{-1}(t_d)} \val(Q(t')) = \mathbb{N},
\]
because $t^{(\tilde{v}\gets \nu)} \in \tr^{-1}(t_d)$ and $\nu \in \val(Q(t^{(\tilde{v}\gets \nu)}))$
for any $\nu\in \mathbb{N}$.
However, there is no $Q'$ satisfying $\val(Q'(t_d)) = \mathbb{N}$ because
the left-hand side is a finite set.
Thus, there exists no~$\RQ~Q$ that satisfies Equation~\forref{qp}.
Conversely, if such $t$~and~$\widetilde{v}$ do not exist,
we can specify the position of~$\tr(t)$ that has the data value queried by~$Q'$ on~$\tr(t)$,
by constructing a tree automaton that simulates each accepting run~$m\in\RUN(A,t^-)$ on~$\tr(t)$.

\subsubsection*{Algorithm \texorpdfstring{$\WQP$}{1-WQP} to Decide Weak Query Preservation}\seclab{1-WQP}
\Stepref{WQP1} constructs a TA~$A_T$ that specifies positions of $t$ that will be deleted by a transducer~$\tr$. \Stepref{WQP2} constructs a product automaton $A'$ that satisfies $L(A') = L(A) \cap L(A_T)$ to find positions that will be in $Q(t)$ and be deleted by $\tr$.

\vspace{2mm}
\noindent
{\bfseries Input:}
$\RQ\ Q=(A,S)$ where $A = (P_{\!A}, \Sigma,P_{\!A}^I, \delta_{\!A})$ is a TA
and $S\,\subseteq\, P_{\!A}$, $\DLTT^V\ \tr = (P_T,$ $\Sigma$, $\Delta$, $\{p_T^0\}$, $\delta_T)$.

\vspace{2mm}
\noindent
{\bfseries Output:}
If $\tr$ weakly preserves~$Q$, output ``Yes,'' otherwise ``No.''

\vspace{2mm}
\noindent
{\bfseries \addStepWQP[\steplab{WQP1}]}
Construct the following TA~$A_T = (P_T \cup \{{\perp}\}, \Sigma,\allowbreak \{p_T^0\}, \delta'_T)$ from~$\tr$ where ${\perp}\notin P_T$ and $\delta'_T$ is the smallest set satisfying the following conditions.
\begin{itemize}
 \item Let $p (\sigma^{(z)} (x_1,\dots,x_n)) \to C^{(j\gets z)}[\,p_1(x_1),
       \dots,p_n(x_n)\,] \in \delta_T$ with $p,p_1,\dots,p_n \in P_T$,
       $\sigma\in\Sigma_n$, and $C \in \ctx(\Delta,\vset_n)$.
       For each~$i\in [n]$, define $\widetilde{p}_i$ as follows.
       If $C$ contains~$x_i$, let~$\widetilde{p}_i = p_i$.
       If $C$ does not contain~$x_i$, let $\widetilde{p}_i = {\perp}$.
       Then, $p \to \sigma(\widetilde{p}_1,\dots,\widetilde{p}_n) \in \delta'_T$.
 \item For each~$\sigma \in \Sigma$, ${\perp} \to \sigma({\perp},\dots,{\perp}) \in \delta'_T$.
\end{itemize}

\vspace{2mm}
\noindent
{\bfseries \addStepWQP[\steplab{WQP2}]}
Construct a product TA~$A'$ of $A$~and~$A_T$ that satisfies
$L(A') = L(A) \cap L(A_T)$. More specifically,
construct the following tree automaton $A'=(P_{\!A}\times P'_T, \Sigma,P_{\!A}^I\times
P_T^I, \delta')$ from~$Q=(A,S)$ and $A_T = (P'_T, \Sigma, P_T^I, \delta'_T)$:
$(p_{\!A}, p_T) \to
\sigma \left( (p_{\!A}^1,p_T^1), \dots, (p_{\!A}^n,p_T^n) \right) \in \delta'$
if and only if $p_{\!A}\to\sigma(p_{\!A}^1,\dots,p_{\!A}^n)\allowbreak \in \delta_{\!A}$ and
$p_T\to\sigma(p_T^1,\dots,p_T^n)\in\delta'_T$.

\vspace{2mm}
\noindent
{\bfseries \addStepWQP[\steplab{WQP3}]}
Remove useless states and rules in~$A'$.
Let $A''=(P'',\Sigma,P''_I,\delta'')$ be the resulting TA.

\vspace{2mm}
\noindent
{\bfseries \addStepWQP[\steplab{WQP4}]}
If the following subset $\PD\subseteq P''$ is empty, output ``Yes,'' otherwise ``No.''
\begin{align*}
\PD = \{\,&
(p,p_T)\in P''\mid p \in S \text{ and (}  p_T = {\perp}
\text{, or}\\
& \text{there are a rule} (p,p_T)\to \sigma((p_{\!A}^1,p_T^1),\ldots,(p_{\!A}^n,p_T^n))\in \delta''\\
& \text{and a value-erasing rule in~$\delta_T$ that has $p_T$~and~$\sigma$ in its left-hand side)} \,\} \qed
\end{align*}

\noindent
We now give a lemma for correctness of our algorithm.

\vspace{3mm}
\begin{lemma}\lmmlab{RQ}
Let $Q$ be a $\RQ$ and $\tr$ be a~$\DLTT^V$.
$\tr$ weakly preserves~$Q$ if and only if
$\PD = \emptyset$ in $\stepref{WQP4}$ of the algorithm~$\WQP$.
\end{lemma}

\subsubsection{Construction of Queries}
If a transducer~$\tr$ weakly preserves a query~$Q$,
a query $Q'=(A',S')$ on target documents
can be constructed by
a type-inference algorithm. The algorithm (called~$\WQC$) works as follows:
(1)~Construct an automaton~$A'$ from~$A$ such that $L(A')=T(L(A))$ where
$T$ is the $\DLTT$ obtained from~$\tr$ by removing the manipulation of values, and
(2)~construct $S'$ accordingly.

\subsubsection*{Algorithm \texorpdfstring{$\WQC$}{1-WQC} to Construct Queries on Target Documents}
\noindent
{\bfseries Input:}
$\RQ~Q=(A,S)$ where $A = (P_{\!A}, \Sigma, P_{\!A}^I, \delta_{\!A})$,
$S\,\subseteq\, P_{\!A}$, $\DLTT^V~\tr = (P_T, \Sigma, \Delta, \{p_T^0\}, \delta_T)$.

\vspace{2mm}
\noindent
{\bfseries Output:} A~$\RQ\ Q'=(A'',S'')$ on target documents that satisfies Equation~\forref{qp}.

\vspace{2mm}
\noindent
{\bfseries \addStepWQC[\steplab{WQC1}]}
Construct a TA $A' = (P_{\!A} \times P_T, \Delta$, $P_{\!A}^I \times \{p_T^0\}$, $\delta')$
from~$Q$ and~$\tr$ where $\delta'$ is defined as follows:
For any rules $r_{\!A}=(p_{\!A} \to \sigma(p_{\!A}^1,\dots,p_{\!A}^n)) \in
\delta_{\!A}$ and
$r_T=(p_T(\sigma^{(z)} (x_1,\dots,x_n)) \to C^{(j\gets
z)}[\,p_T^1(x_1),\dots,p_T^n(x_n)\,]) \in\delta_T$,
\begin{itemize}
 \item For each $v\in \pos(C)$ such that $\lab(C,v)\in \Delta$,
       \[
       M(v) \to \lab(C,v)(M(v1),\dots,M(vn_v))\in \delta'
       \]
       where $n_v=\rk(\lab(C,v))$ and $M$ is a mapping such that for each
       $v\in \pos(C,v)$,
 \begin{itemize}
  \item $M(v)=(p_{\!A},p_T)$ if $v=\varepsilon$,
  \item $M(v)=(p_{\!A}^i,p_T^i)$ if $\lab(C,v)=x_i\in \vset_n$, and
  \item $M(v)=(r_{\!A},r_T,v)$ otherwise;
 \end{itemize}
 \item $(p_{\!A}, p_T) \to (p_{\!A}^i,p_T^i) \in \delta'$ if $C=x_i\in \vset_n$.
\end{itemize}

\noindent
{\bfseries \addStepWQC[\steplab{WQC11}]}
Construct a reduced TA without epsilon rules equivalent with $A'$.
Formally, let $\tilde{p}\Rightarrow_\varepsilon \tilde{p}'$ if and only if
$\tilde{p}\rightarrow \tilde{p}'\in \delta'$,
and $\Rightarrow^*_\varepsilon$ be the reflexive transitive closure of $\Rightarrow_\varepsilon$.
For each rule $\tilde{p}\rightarrow
\sigma(\tilde{p}_1,\ldots,\tilde{p}_n)\in \delta'$
and $\tilde{p}'_1,\ldots,\tilde{p}'_n\in \stat(A')$,
add to $\delta'$ a new rule
$\tilde{p}\rightarrow \sigma(\tilde{p}'_1,\ldots,\tilde{p}'_n)$
if for $i\in [n]$,
$\tilde{p}_i\Rightarrow^*_\varepsilon \tilde{p}'_i$ and
there is a rule with $\tilde{p}'_i$ in its left-hand side and
some symbol in $\Delta$ in its right-hand side.
Then, remove all epsilon rules, useless states and transition rules of $A'$.
Let $A''$ be the resulting TA.

\vspace{2mm}
\noindent
{\bfseries \addStepWQC[\steplab{WQC2}]}
Compute $S''=\bigcup_{i=1}^n S_{p_i}$ where $S_{p_i}$ is the smallest subset of
$\stat(A'')$ satisfying the following conditions.
\begin{itemize}
 \item $(r_{\!A},r_T,v)\in S_{p_i}$ if
       $r_{\!A}$ has $p_i$ in its left-hand side, and
       the right-hand side of $r_T$ is
       \[
        C^{(j\gets z)}[\,p_T^1(x_1), \dots,p_T^n(x_n)\,])
       \]
       where $j\neq \varepsilon$ and $v=j$.
 \item $(p_i,p_T)\in S_{p_i}$ if $p_T(\sigma^{(z)} (x_1,\dots,x_n)) \to C^{(j\gets z)}
       [\,p_T^1(x_1),\dots,p_T^n(x_n)\,]) \in\delta_T$ where
       $j=\varepsilon$.\qed
\end{itemize}

\vspace{3mm}
\begin{theorem}\thmlab{RQ}
 Given a~$\RQ~Q$ and a $\DLTT^V~\tr$,
 it is decidable whether $\tr$ weakly preserves~$Q$ or not.
 Furthermore, if it preserves, the query~$Q'$ satisfying Equation~$\forref{qp}$
 can be constructed by~$\WQC$ in PTIME\null.
\end{theorem}

\subsection{General Case}
\label{sec:general case}
 If the tuple size~$n$ of queries is constant,
 we can solve the weak query preservation problem for~$\RQN$ under~$\DLTT^V$ in polynomial time by using the algorithm for unary queries.
We sketch an algorithm for general case below. Let $Q=(A,S)$ be a given~$\RQN$.
We can see that a $\DLTT^V~\tr$ weakly preserves~$Q$ if and only if
for every~$s\in S$, $\tr$ weakly preserves~$Q_s = (A,\{s\})$.
Also $Q(t) = \bigcup_{s\in S} Q_s(t)$.
Thus we will assume that $\left|S\right|=1$ and let $s = (p_1, \dots, p_n)$.
The basic idea is to consider $Q' = (A, \{p_1,\dots,p_n\})$ instead of~$Q$
and test whether $\tr$ weakly preserves $Q'$.
However, this does not work in general because
$Q(t)$ contains only a tuple~$(v_1, \dots, v_n)$ of positions such that
there is a run~$m\in \RUN(A,t^-)$ satisfying $m(v_i)=p_i$ for each~$i$ ($1\le i\le n$)
simultaneously while $Q'(t)$ contains every position~$v$
such that there is a run~$m\in \RUN(A,t^-)$ satisfying $m(v)=p_i$
even if there is some~$p_j$ ($j\neq i$) such that for any~$u\in\pos(t)$, $m(u)\neq p_j$.

\begin{example}
 Let $Q = (A_s, S)$ be $3\mathchar`-\mathrm{RQ}$ defined by
 \begin{align*}
  A_s &= (P, \Sigma, P_I, \delta),\ P = \{p_1,\, p_2,\, p_3,\, p_\# \},
  \ \Sigma = \Sigma_2\cup\{\#\},\ \Sigma_2 = \{A, B, C\},\ P_I = \{ p_1 \}, \\
  \delta &= \{\, p_1 \to A(p_2, p_\#),\ \ p_2 \to B(p_3, p_\#),
   \ p_2 \to C(p_\#, p_\#),\ p_3 \to C(p_\#, p_\#),\ p_\# \to \#\,\},\\
  S &= \{(p_1, p_2, p_3)\}.
 \end{align*}
Also let $\tr=(P, \Sigma, \Sigma, \{p_1\}, \delta_T)$ be $\DLTT^V$ defined by
\begin{align*}
 \delta_T = \{\,&p_1(A^{(z)}(x_1,x_2)) \to A^{(\varepsilon \gets z)}(p_2(x_1), p_\#(x_2)), \\
 & p_2(B^{(z)}(x_1, x_2)) \to B^{(\varepsilon \gets z)}(p_3(x_1), p_\#(x_2)), \\
 & p_2(C^{(z)}(x_1, x_2)) \to \#, \\
 & p_3(C^{(z)}(x_1, x_2)) \to C^{(\varepsilon \gets z)}(p_\#(x_1), p_\#(x_2)),\\
 & p_\# (\#) \to \#\,\},
\end{align*}
 where $P$ and $\Sigma_2$ are the same as~$A_s$. Consider the following data trees:
\begin{align*}
 t_1 &= A^{(1)}(B^{(2)}(C^{(3)}(\#,\#),\#),\#),\\
 t_2 &= A^{(1)}(C^{(3)}(\#,\#),\#).
\end{align*}

\figref{proof} shows $Q(t_1)$, $Q(t_2)$, $\tr(t_1)$ and $\tr(t_2)$, the application results of
$Q$ and $\tr$ to $t_1$ and $t_2$.
In fact, for any data tree~$t$, TA~$A_s$ never assigns $p_3$ to any position of $t$
 (and thus $Q(t) = \emptyset$ because $(p_1, p_2, p_3)$ contains $p_3$)
 if and only if
  $\DLTT^V\ \tr$ deletes a subtree of $t$ such that $\tr$ assigns $p_2$ to the root of
the subtree.
That is, the deletion of a subtree by $\tr$ does not violate the
weak query preservation for~$Q$.
However, if we consider $\RQ\ Q'=(A_s, \{p_1,p_2,p_3\})$ instead of $Q$ and
apply $\WQP$ to $Q'$, then $\WQP$ answers ``No''
(because $\tr$ does not weakly preserve $Q'$).

To overcome the above mentioned problem, we modify $A_s$ as $A^{\!F}$ so that
$m\in \RUN(A^{\!F},t^-)$ only if for every~$p_j$ ($1\le j\le n$), there is
$u_j\in\pos(t)$ such that $m(u_j)=p_j$.
This modification can be done by augmenting each state~$p$
with a subset~$P$ of $\{p_1, \dots, p_n\}$.
For $m\in \RUN(A^{\!F},t^-)$, if $m$ assigns $(p,P)$ to a position~$v$,
each state in $P$ should be used at least once as the first component
of a state in the subtree rooted at $v$ (including $v$).
Especially, an initial state of $A^{\!F}$ is a pair of an initial state
of $A_s$ and $\{p_1,\dots,p_n\}$, meaning that each~$p_i$
($1\le i\le n$) should be used at least once in the input tree.
$A^{\!F}$ is weakly preserved by~$\tr$ if and only if
the original~$A_s$ is weakly preserved by~$\tr$.

 \begin{figure}[t]
  \begin{center}
   \includegraphics[width=400pt]{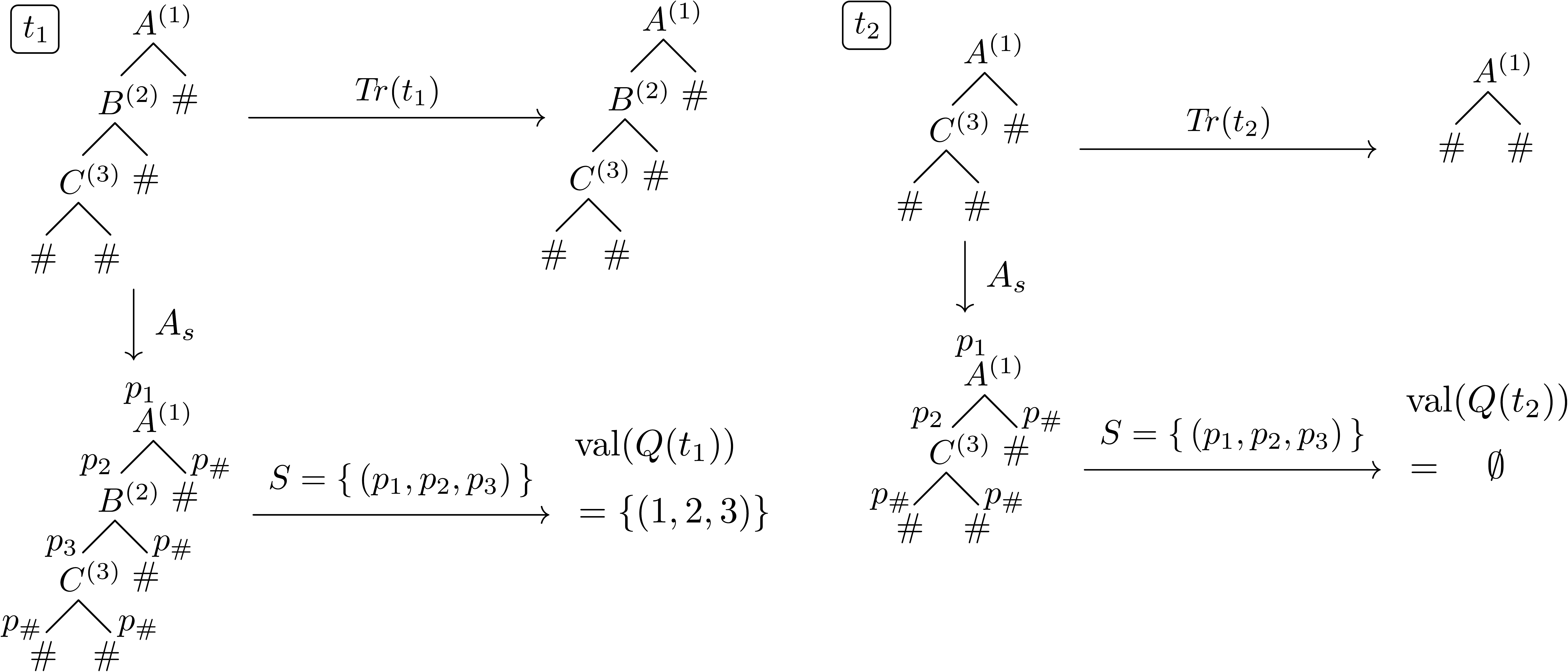}
  \end{center}
  \vspace{-3mm}
  \caption{Application results of $Q$ and $\tr$.}
  \figlab{proof}
 \end{figure}

\end{example}

\subsubsection*{Algorithm \texorpdfstring{$\WQPN$}{n-WQP} to Decide Weak Query Preservation}\seclab{n-WQP}
{\bfseries Input:}
$\RQN\ Q = (A, S)$, $A = (P_{\!A}, \Sigma,P_{\!A}^I, \delta_{\!A}),\ S\,\subseteq\, P_{\!A}^n$,
$\DLTT^V\ \tr = (P_T, \Sigma, \Delta, \{p_T^0\}, \delta_T)$.
We can assume that the set~$S$ of tuples of states has just one element~$(p_1,\dots,p_n)$, i.e., $S=\{(p_1,\dots,p_n)\}$, by the reason described above.

\vspace{2mm}
\noindent
{\bfseries Output:}
If $\tr$ weakly preserves~$Q$, output ``Yes,'' otherwise ``No.''

\vspace{2mm}
\noindent
{\bfseries \addStepWQPN[\steplab{WQPN1}]}
Let $P_{\!s}=\{p_1,\ldots,p_n\}\subseteq P_{\!A}$.
Construct $A^{\!F}=(P_{\!A}\times 2^{P_{\!s}}, \Sigma, P^I_{\!A}\times \{P_{\!s}\}, \delta^F)$ from $A$
where $\delta^F$ is defined as follows:
For $p \to a(p^1_{\!A},\dots,p^m_{\!A})\in \delta_{\!A}$ and $P, P_1,\dots,P_m\subseteq P_{\!s}$,
\[
(p,P) \to a((p^1_{\!A},P_1),\dots,(p^m_{\!A},P_m)) \in \delta^F
\]
if and only if for each $i\in [n]$,
\begin{itemize}
 \item if $p_i\in P$ and $p=p_i$, then $p_i\notin P_j$ for all $j\in [m]$,
 \item if $p_i\in P$ and $p\neq p_i$, then there exists exactly one $j\in [m]$ satisfying $p_i\in P_j$, and
 \item if $p_i\notin P$, then $p_i\notin P_j$ for all $j\in [m]$.
\end{itemize}

\vspace{2mm}
\noindent
{\bfseries \addStepWQPN[\steplab{WQPN2}]}
Let $Q^{\!F}=(A^{\!F}, \{p_1,\dots,p_n\}\times 2^{P_{\!s}})$.
Decide whether $\tr$ weakly preserves $Q^{\!F}$ by~$\WQP$. \qed

We have that $\tr$ weakly preserves $Q$ if and only if
$\tr$ weakly preserves $Q^{\!F}$.

If $n$ is not fixed, however, the above algorithm will take exponential time in~$n$.
The following theorem gives a lower bound for the weak query preservation problem.

\vspace{1mm}
\begin{theorem}
\label{th:RQN}
 Given an~$\RQN~Q$ and a~$\DLTT^V~\tr$,
 the problem of deciding whether $\tr$ weakly preserves~$Q$ is coNP-complete.
 If it preserves, a query~$Q'$ satisfying Equation~$\forref{qp}$ can be constructed.
 \end{theorem}
\vspace{1mm}

We can create $n$-ary queries on the transformed documents by using a
natural variant of $\WQC$.
Given a $\DLTT^V~\tr$ and $\RQ~Q=(A,\{(p_1,\dots,p_n)\})$,
the variant (called $\WQCN$) works in almost the same way as $\WQC$
except that it computes $S''=\prod_{i=1}^n S_{p_i}$ at~\stepref{WQC2}
while $\WQC$ computes $S''=\bigcup_{i=1}^n S_{p_i}$.

\section{Decidability of Query Preservation}
\seclab{SQP}
We will provide an algorithm~$\QPN$ that decides query preservation.
To begin with, we give the following lemma.

\vspace{1mm}
\begin{lemma}
\label{lm:nf}
Let $Q=(A,\{s\})$ be an~$\RQN$.
We can construct an $\RQN$ $\widetilde{Q}=(\widetilde{A},\widetilde{S})$
equivalent with $Q$ such that
$(1)$~there exist pairwise disjoint subsets $S_1,\dots,S_n$ of $\stat(\widetilde{A})$
satisfying $\widetilde{S}=S_1\times\dots\times S_n$,
and
$(2)$~for every~$t^-\in L(\widetilde{A})$, $m\in\RUN(\widetilde{A}, t^-)$, and $i$ $(1\le i\le n)$,
there exists exactly one~$v\in\pos(t)$ such that $m(v)\in S_i$.
\end{lemma}
\vspace{1mm}
Thus, any $\RQN$ $Q=(A,\{s_1,\dots,s_k\})$ can be represented as the union of $\RQN$s $\widetilde{Q}^1,\dots,\widetilde{Q}^k$
such that each $\widetilde{Q}^j$ $(j\in [k])$ is equivalent with $(A,\{s_j\})$ and satisfies the two conditions in Lemma~\ref{lm:nf}.


We explain the idea of our algorithm~$\QPN$.
Assume we are given a~$\DLTT^V~\tr$ and an $\RQN~Q=(A,S)$ where $S=S_1\times \cdots\times S_n$
satisfying the conditions (1) and (2) in Lemma~\ref{lm:nf}.
As shown in Example~\ref{ex:1},
if $\tr$ weakly preserves $Q$ but $\tr$ does not strongly preserve $Q$, there are
two different data trees $t_1$ and $t_2$ such that $\tr(t_1)=\tr(t_2)(=t')$,
runs $m_1\in\RUN(A,t^-_1)$, $m_2\in\RUN(A,t^-_2)$,
$m_1(v_1)$, $m_2(v_2)\in S_i$,
positions~$v_1\in\pos(t_1)$, and $v_2\in\pos(t_2)$ that have different values.

To test whether this situation happens,
the algorithm~$\QPN$ introduces a ``marked'' input symbol~$(i,\sigma)$
for each $i\in [n]$ and $\sigma\in\Sigma$;
$\QPN$ first constructs a TA~$A_\mk$ that simulates~$A$
except that for each~$i\in [n]$
if $A$ assigns a state in $S_i$ when reading $\sigma$
at some position~$v$ (which is unique by the conditions in Lemma~\ref{lm:nf}),
$A_\mk$ assigns a state in $S_i$ when reading $(i,\sigma)$ at~$v$.
Then, $\QPN$ constructs $A'_\mk$ such that
$L(A'_\mk) = T_\mk^{-1}(T_\mk(L(A_\mk)))$, where $T_\mk$ is a ``marked'' version of $\tr$.
This construction is possible
because $\LTT^V$ and its inverse preserve regularity.
Also, it constructs $A_\idx$, which simulates $A$
while identifying a marked symbol~$(i,\sigma)$ with~$\sigma$.
By the reason stated above, we can prove that
$\tr$ preserves~$Q$ if and only if
$L(A_\mk) = L(A_\idx) \cap L(A'_\mk)$.

\subsubsection*{Algorithm \texorpdfstring{$\QPN$}{n-QP} to Decide Query Preservation}\seclab{n-SQP}
\noindent {\bfseries Input:}
$\RQN$~$Q=(A,\{s_1,\dots,s_k\})$, $\DLTT^V~\tr= (P_T, \Sigma,\allowbreak \Delta, \allowbreak\{p_T^0\}, \delta_T)$.
We assume that
$\tr$ weakly preserves~$Q$ (which is a necessary condition that can be decided as shown in Theorem~\ref{th:RQN}).

\vspace{2mm}
\noindent
{\bfseries Output:}
If $\tr$ preserves~$Q$, output ``Yes,'' otherwise ``No.''

\vspace{2mm}
\noindent
{\bfseries \addStepQP[\steplab{SNQP0}]}
For each~$s_j$ $(j\in [k])$, construct $\widetilde{Q}^j$ equivalent with $(A,\{s_j\})$ satisfying the conditions (1) and (2) in Lemma~\ref{lm:nf}.

\vspace{2mm}
\noindent
{\bfseries \addStepQP[\steplab{SNQP1}]}
For each $\widetilde{Q}^j=(A^j,S^j)$ where $A^j=(P^j,\Sigma, P_{I}^j,\delta^j)$ and $S^j=S^j_1\times\cdots\times S^j_n$,
construct TA~$A^j_\mk=(P^j, \Sigma\cup([n]\times\Sigma), P_{I}^j, \delta_{\!Aj}^\mk)$
from~$A^j$ where $\delta_{\!Aj}^\mk$ is defined as follows.
For each rule of the form
$p\to \sigma(p_1,\dots,p_m) \in \delta^j$, if $p$ is in $S^j_i$
then $p\to (i,\sigma)(p_1,\dots,p_m) \in \delta_{\!Aj}^\mk$,
otherwise $p\to \sigma(p_1,\dots,p_m) \in \delta_{\!Aj}^\mk$.
Then, construct TA~$A_\mk$ as the union TA of $A^1_\mk, \ldots, A^n_\mk$.

\vspace{2mm}
\noindent
{\bfseries \addStepQP[\steplab{SNQP2}]}
Construct $\DLTT~T_{\!\mk}=
(P_T, \Sigma\cup([n \steplab{NQP1}]\times\Sigma),
\Delta\cup([n]\times\Delta), \{p_T^0\}, \delta_T^\mk)$ where
$\delta_T^\mk$ is the smallest set satisfying the following conditions:
For each~$i\in [n]$ and for each rule $p_T(\sigma^{(z)}(x_1,\dots,x_m))
\to C^{(j\leftarrow z)}[\,p_T^1(x_1),\dots,p_T^m(x_m)\,] \in \delta_T$,
let
\[
 p_T(\sigma(x_1,\dots,x_m))
  \to C[\,p_T^1(x_1),\dots,p_T^m(x_m)\,] \in \delta_T^\mk,
\]
and
\[
 p_T((i,\sigma)(x_1,\dots,x_m))
  \to \widebar{C}[\,p_T^1(x_1),\dots,p_T^m(x_m)\,] \in \delta_T^\mk
\]
where $\lab(\widebar{C},j)=(i,\lab(C,j))$, and for each~$v\in\pos(C)$
satisfying $v\neq j$, $\lab(\widebar{C},j)=\lab(C,j)$.

\vspace{2mm}
\noindent
{\bfseries \addStepQP[\steplab{SNQP3}]}
Construct TA~$A'_\mk$ such that
\[
 L(A'_\mk)=T_\mk^{-1}(T_\mk(L(A_\mk)))
\]
by type inference and inverse type inference.

\vspace{2mm}
\noindent
{\bfseries \addStepQP[\steplab{SNQP5}]}
Construct TA~$A_\idx=(P,\Sigma\cup([n]\times\Sigma), P_I\cup([n]\times P_I),\delta\cup\delta_\idx)$
from~$A$ where $\delta_\idx$ is defined as follows. For each rule of the form
$p\to \sigma(p_{\!A}^1,\dots,p_{\!A}^m) \in \delta$ and for each~\mbox{$i\in[n]$},
let $p\to (i,\sigma)(p_{\!A}^1,\dots,p_{\!A}^m) \in \delta_\idx$.

\vspace{2mm}
\noindent
{\bfseries \addStepQP[\steplab{SNQP6}]}
If $L(A_\mk) = L(A_\idx) \cap L(A'_\mk)$, output ``Yes,'' otherwise ``No.''\qed

\begin{example}
\label{ex:2}
Recall $Q$~and~$\tr$ in Example~\ref{ex:1}.
By steps~1--4 of the above algorithm~$\QPN$,
\begin{align*}
L(A_\mk) & = \{f((1,a),a), g(a,(1,a))\}, \\
L(A'_\mk) & = \{f((1,a),a), g((1,a),a), f(a,(1,a)), g(a,(1,a))\}\\
 & = L(A_\idx) \cap L(A'_\mk).
\end{align*}
Hence, $L(A_\mk) \varsubsetneq L(A_\idx) \cap L(A'_\mk)$ holds
and the algorithm answers ``No.''
\end{example}

\vspace{2mm}
\begin{lemma}\lmmlab{SNQP}
 If $\DLTT^V~\tr$ weakly preserves~$\RQN~Q$,
 then $\tr$ preserves~$Q$ if and only if $L(A_\mk) = L(A_\idx) \cap L(A'_\mk)$
 in $\stepref{SNQP5}$ of the algorithm~$\QPN$.
\end{lemma}

\vspace{2mm}
\begin{theorem}\thmlab{SRQN}
 Given an~$\RQN~Q$ and a $\DLTT^V~\tr$,
 the problem of deciding whether $\tr$ preserves~$Q$ is in~2-EXPTIME\null.
 If it preserves, a query~$Q'$ constructed by~$\WQCN$ satisfies Equation~$\forref{sqp}$.
\end{theorem}
\vspace{2mm}

\section{Conclusion}
We have studied the decidability problems of the weak query preservation
and the strong query preservation.
We have modeled an XML document by a data tree,
a document transformation by a deterministic linear top-down data tree transducer,
and a query to the tree by a run-based $n$-ary query.
We showed the weak query preservation problem is coNP-complete for $n$-ary queries
where $n$ is not fixed, and the problem becomes PTIME if $n$ is a constant.
We also showed the strong query preservation problem is in 2-EXPTIME\null.

\subsection*{Future Work}
Our model of tree transformations does not allow copying of elements on trees.
Copying elements is one of the fundamental operations for XML, so it would be important
to know whether the query preservation problems are decidable or not
for a transformation model having a copy operation.

\subsection*{Acknowledgements}
We would like to thank two anonymous referees
for their helpful comments, which greatly improved the paper.

%
\label{sect:bib}

\end{document}